# Intraday probabilistic forecasts of solar resources with cloud scale-dependent autoregressive advection

Carpentieri, A.[a,b], S. Pulkkinen[c], D. Nerini[d], D. Folini[b], M. Wild[b], A. Meyer[a]

[a] Bern University of Applied Sciences, Quellgasse 21, 2501 Biel, Switzerland
[b] Institute for Atmospheric and Climate Science, ETH Zurich, Universitaetstrasse 16, 8092 Zurich, Switzerland
[c] Finnish Meteorological Institute, PL 503, FIN-00101 Helsinki, Finland
[d] Federal Office of Meteorology and Climatology MeteoSwiss, Locarno-Monti, Switzerland
Corresponding author: alberto.carpentieri@bfh.ch

**Abstract**. Solar energy supply is usually highly volatile which limits its integration in the power grid. Accurate probabilistic intraday forecasts of solar resources are essential to increase the share of photovoltaic (PV) energy in the grid and enable cost-efficient balancing of power demand and supply. Solar PV production mainly depends on downwelling surface solar radiation (SSR). By estimating SSR from geostationary satellites, we can cover large areas with high spatial and temporal resolutions, allowing us to track cloud motion. State-of-the-art probabilistic forecasts of solar resources from satellite imagery account only for the advective motion of clouds. They do not consider the evolution of clouds over time, their growth, and dissipation, even though these are major sources of forecast uncertainty. To address the uncertainty of cloudiness evolution, we present SolarSTEPS, the first optical-flow probabilistic model able to simulate the temporal variability of cloudiness. We demonstrate that forecasting the autocorrelated scale-dependent evolution of cloudiness outperforms state-of-the-art probabilistic advection-based forecasts by 9% in continuous ranked probability score (CRPS). This corresponds to an extension of the forecast lead time by about 45 minutes at constant CRPS. Our work is motivated by the scale-dependent predictability of cloud growth and decay. We demonstrate that cloudiness is more variable in time at smaller spatial scales than at larger ones. Specifically, the temporal autocorrelation of cloudiness is related to its spatial scale by a rational function. We also demonstrate that decomposing cloudiness into multiple spatial scales in the forecasts further improves the forecast skill, reducing the CRPS by 10% and the RMSE by 9%. By exploiting more than one autoregressive model, we can simulate the chaotic evolution of small scales independently of the predictable evolution of larger spatial scales.

**Key points:**
- Probabilistic forecast model with scale-dependent cloudiness variability.
- We show that the temporal evolution of cloudiness depends on spatial scale.
- Predicting the cloud evolution variability separately for different scales improves the forecast quality.
- Our model SolarSTEPS outperforms probabilistic advection with an 9.3% improved CRPS.
- SolarSTEPS extends solar forecasts by about 45 minutes at the same forecast skill.
- Using a local covariance matrix in the ensemble generation improves the forecast skill.

# 1. Introduction
Solar radiation is an abundant source of renewable energy. Harvesting it to replace fossil fuels is a central pillar of climate change mitigation. The installed solar power capacity has continuously increased in Europe and globally in the past years and is expected to grow even more strongly in the coming decades (Bojek et al., 2021). But surface solar radiation (SSR) highly depends on local weather and, in particular, the presence, motion, and optical thickness of clouds. The resulting variability of SSR over the course of a day can cause highly fluctuating solar power production. Therefore, increasing shares of solar energy in the power grid can confront grid operators with challenging balancing tasks.

Accurate intra-day forecasts of SSR are essential for the informed decision-making of power grid operators, utility companies, and energy traders. Likewise, it is also important to characterize the spread and uncertainty of the SSR forecasts so that the decision-making processes can be supported with accurate risk estimates on the future SSR values. In fact, deterministic forecasts cannot be reliable as they cannot distinguish between noise and signal. Assuming uniformly distributed measurements, a deterministic forecast would be useless. On the other hand, estimating the probability distribution of future values can characterize noisy/unpredictable situations. Thus, well-characterized forecast uncertainties facilitate reductions in the stand-by capacity and cost of flexible conventional power generators (spinning reserves) and energy storage required to balance the grid (Kaur et al., 2016). David et al., 2021 found that a 1% rise in the forecast accuracy brings about a 2% increase in economic profit in a case study of a multi-Megawatt PV plant in Australia with Lithium-ion battery storage. The finding illustrates that even a small gain in forecast accuracy on the level of a few percent is of potential economic interest.

A vast literature exists on short-term forecasting of surface solar radiation (SSR). Comprehensive reviews of SSR forecasting methods have been provided by Yang et al., 2022; Sobri et al., 2018; Van der Meer et al., 2018; and Inman et al., 2013. SSR forecasting approaches either involve physics-based models that rely on meteorological input data for their initialization, or statistical and machine learning models that use other input data types or no exogenous input data at all. Physics-based forecasting approaches usually make use of numerical weather prediction models or observation systems of clouds and aerosols. These observation systems include satellite-based imagers (Gallucci et al., 2018; Arbizu-Barrena et al., 2017; Schroedter-Homscheidt et al., 2016), pyranometer networks (Chow et al., 2011), sky cameras (Rodriguez-Benitez et al., 2021; Cheng, 2017; Kuhn et al., 2017a, 2018a; Kazantzidis et al., 2017; Alonso-Montesinos et al., 2015) and ground-facing shadow cameras (Kuhn et al., 2017b, 2018b). Statistical and ML-based forecasting models that do not make use of any meteorological initialization data have been proposed as well. These models usually capture the autocorrelated nature of cloudiness time series at a certain location and require historical SSR data for the model training (Pacurar et al., 2013; Pedro et al., 2012).

Deterministic short-term forecasting methods of SSR cover a large range of spatial scales. Satellite-based SSR models enable forecasts at regional to intercontinental scales due to the satellites' wide field of view. Geostationary satellites capture images of the Earth several times per hour, allowing continuous monitoring of any region of interest between 65°N and 65°S. In contrast, satellites in low Earth orbit are less suitable for SSR nowcasting applications because of their constantly changing ground footprints and few daily overpasses. Sky cameras are used to continuously image the sky at a certain location. Sky cameras can provide cloudiness maps at high temporal and spatial resolution. However, their spatial coverage is only on the scale of kilometers. High spatio-temporal resolution and a sufficient large coverage permit to track the cloud movements by using optical-flow techniques (Aicardi et al., 2022). Cloud-tracking is defined by cloud motion vectors (CMVs). CMV fields describe the velocity and direction for every pixel on the map. They can be derived from both sky cameras and satellite imagers to forecast SSR at short lead times (Aicardi et al., 2022; Arbizu-Barrena et al., 2017; Chow et al., 2011). The Lucas-Kanade (LK) method can be applied to derive cloud motion vector fields (Lucas and Kanade, 1981). It compares successive satellite or sky camera images to identify matching features. CMVs are computed for the features to estimate their future paths. Satellite CMV-based methods have been shown to outperform numerical weather prediction methods for forecasting cloud motion for lead times of up to 4 hours (Wang et al., 2019). Starting from Leese et al., 1971 by using a cross-correlation method to detect movements, many works in the literature exploited satellite maps to derive cloud motion vectors. In Hammer et al., 1999, a probabilistic approach based on Konrad et al., 1992 was used to forecast Meteosat-based cloud index maps. In Perez et al., 2010, different operational CMV methods are compared to forecast clear-sky index maps, and the best performing method is selected for each pixel. More recently, similar approaches have been used to forecast cloud properties such as cloud height and thickness (Batlles et al., 2014, Wang et al., 2019), while in Gallucci et al., 2018, the authors computed the advection field directly on the SEVIRI different spectral channel maps and then converted them into solar radiation with a radiative transfer model. However, the most common features used to extract the advection fields are the cloudiness indices (Urbich et al., 2018). Finally, in Aicardi et al., 2022, five different optical-flow algorithms are tested and compared to forecast hourly cloudiness maps, and the results showed that TV-L1 and hierarchical Lucas-Kanade are the two best-performing methods.

State-of-the-art probabilistic short-term SSR forecast models are still based on numerical weather forecasts or historical in-situ SSR point measurements rather than on near-real-time observations of the plant surrounding area (David et al., 2018). This is surprising because forecast methods using satellite data and CMVs hold great potential for higher accuracy intraday forecasts with well-characterized uncertainties (Yang et al., 2018). Several satellite-CMV-based SSR forecasting methods have been demonstrated in the past decade (e.g., Wang et al. 2019; Urbich et al., 2018, Blanc et al., 2017; Coimbra et al., 2013) but they provide deterministic forecasts that lack the capability to characterize forecast uncertainties.

Probabilistic methods for CMV-based SSR forecasting have barely been developed up to now. Recently, Carriere et al., 2021 presented the first probabilistic satellite- and CMV-based SSR forecast model. The authors presented a probabilistic cloud advection approach that adds perturbations to the advection field to create a forecast ensemble. However, the model does not capture the evolution of clouds over time, their growth and dissipation, even though cloud evolution is a major source of forecast uncertainty. Clouds are the results of complex dynamical, thermodynamical, and chemical interactions (Lohmann et al., 2016). A purely advective forecast approach cannot capture these processes. In addition, the proposed model forecasts SSR only at single sites, so its forecasts are barely suited for estimating regionally aggregated solar resources and PV production because single sites tend to lack representativeness of their surrounding regions (Li et al., 2005; Hakuba et al., 2013; Li, 2014; Huang et al., 2016; Schwarz et al., 2018).

We present a probabilistic satellite-based approach for intraday forecasts of SSR that accounts for the temporal evolution and spatial structure of clouds. Our forecast model captures the autocorrelated nature and size dependent evolution of cloudiness in addition to the cloud advection. Our forecast model predicts SSR for the entire area covered by the satellite image or parts thereof. So, its forecasts are a suitable basis for applications such as predicting regionally aggregated PV production.

This paper is organized as follows. Section 2 details the proposed probabilistic intraday forecast approach. The dataset, benchmarks, and metrics used for characterizing the forecast quality are described in section 3. Section 4 discusses the quality of our SSR forecast approach. Section 5 presents conclusions and opportunities for future research.

## 2. Intraday probabilistic forecast model SolarSTEPS

The forecast task we address is to predict the probability density distributions of the upcoming $n$ clear-sky index maps based on a sequence of $m$ clear-sky index maps, $(CSI_t)_{t_0-(m-1)}^{t_0}$. The clear-sky index (CSI) corresponds to the ratio of the (all-sky) SSR to the clear-sky SSR. Unlike SSR, the clear-sky index does not suffer from daily or annual seasonalities which makes it a more suitable variable to forecast than SSR. Our aim is to forecast a probability density function (PDF) of the clear-sky index, $p(CSI_t)$, that describes the probabilities associated with the cloudiness at any location $(x, y)$ for $n$ future time steps $t = t_0 + 1, \dots, t_0 + n$. We refer to the cloud radiative effect represented by the clear-sky index as cloudiness. Both terms - clear-sky index and cloudiness - are used interchangeably throughout the text.

### 2.1 Forecast model steps

We introduce a probabilistic satellite-based model for intraday forecasts of SSR, SolarSTEPS, that accounts for the advection of the cloud field as well as for the temporal evolution and spatial structure of cloudiness. Our method is inspired by precipitation nowcasting research (Pulkkinen et al., 2019; Bowler et al., 2006) which investigated scale-dependent autoregressive modeling of precipitation fields to simulate their scale-dependent temporal variability. Forecasts with the SolarSTEPS model involve the following sequence of steps (Fig. 1):

1) First, cloudiness (clear-sky index) maps are estimated from Meteosat SEVIRI Level 1.5 data by the HelioMont algorithm. Specifically, maps of the clear-sky index are derived based on Castelli et al., 2014.
2) Cloud motion vectors are computed from the $m$ most recent maps of the clear-sky index based on the Lucas-Kanade algorithm (Lucas and Kanade, 1981).
3) The clear-sky index maps derived in step 1 are decomposed into $n_c$ cascade levels and corresponding cascade-level maps (Fig. 2) to account for the scale-dependent life time. This is achieved by Fast Fourier transforming (FFT) the maps, and then applying Gaussian bandpass filters to the resulting FFT output. Each cascade level represents a different spatial scale, as illustrated in Fig. 2 and Tab. 1.
4) An autoregressive (AR) model to simulate the cloud evolution is estimated for each cascade level on cascade-level maps in the Lagrangian reference frame using a semi-Lagrangian advection method (see Section 2.2).
5) $k_{ens}$ spatially correlated noise fields are computed to generate an ensemble of forecasts by perturbing the AR models resulting in a probabilistic forecast. After the AR iterations, the cascade levels forecasts are merged together to produce the forecast.
6) Finally, the AR ensemble forecasts are advected using the cloud motion vectors computed in step 2. The PDFs of the clear-sky index are approximated based on the estimated forecast ensemble, $\left\{\widehat{Kl}_t^{\,1}(x,y), \ldots, \widehat{Kl}_t^{\,k_{ens}}(x,y)\right\}$. Eventually, the CSI maps can be converted to SSR with a pixel-wise multiplication by the estimated clear-sky SSR, $SSR_{cs}$.

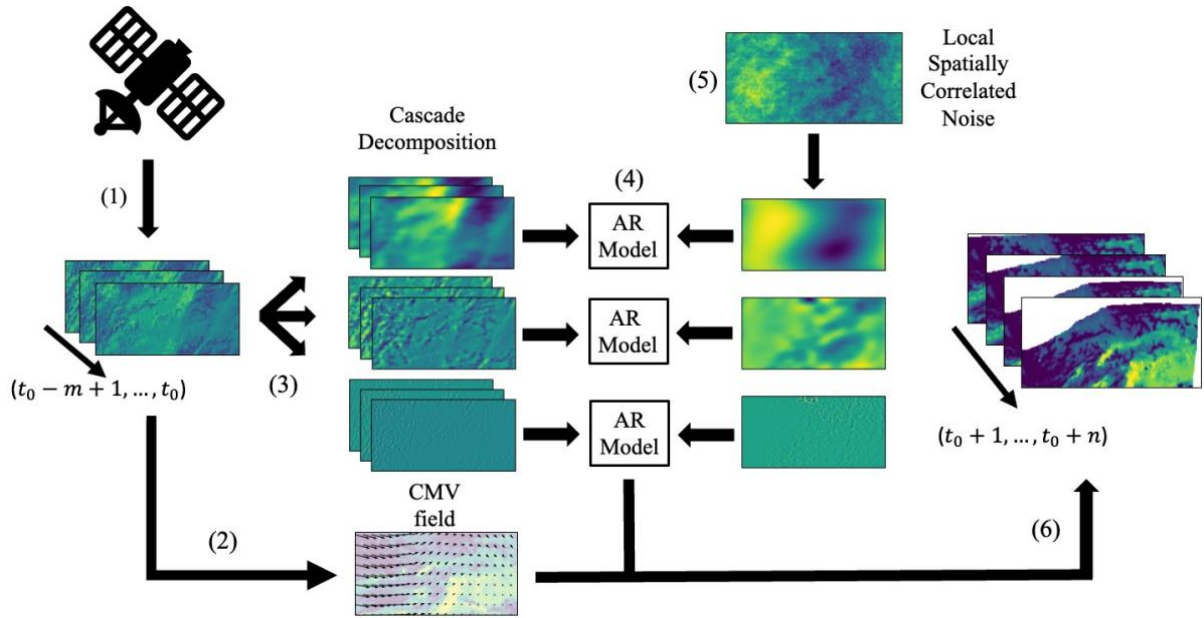

**Figure 1.** The probabilistic solar forecast approach of SolarSTEPS based on clear-sky index maps, shown for a single ensemble member. A forecast with three cascade levels and three associated AR models is shown as an example case in which three past time steps (m=3) are used to forecast four future time steps (n=4), corresponding to forecast lead times of +15 to +60 minutes. Multiple ensemble members are generated (step 5) in our probabilistic forecasts. The figure shows the forecast of only a single ensemble member to improve the readability of the figure.

## 2.2 Cloud advection

The advection of successive clear-sky index maps is modeled using the Lucas-Kanade (LK) optical-flow algorithm (Lucas and Kanade, 1981). It computes the advection field following the assumption of brightness constancy, which is described by:

$$\frac{dKI}{dt} = \frac{\partial KI}{\partial x}\frac{dx}{dt} + \frac{\partial KI}{\partial y}\frac{dy}{dt} + \frac{\partial KI}{\partial t} = 0 \quad (1)$$

Eq. (1) describes the invariance of the pixel intensities of a moving object. The LK algorithm applies it on a local scale by using a moving window. The advection is performed using the semi-Lagrangian method by Germann and Zawadzki, 2002 on the motion field described by the velocities $v_x = \frac{dx}{dt}$ and $v_y = \frac{dy}{dt}$. The version of LK used in this study is the pyramidal implementation by Bouget, 2001, available in the Pysteps (Pulkkinen et al., 2019) and OpenCV (Bradski, 2000) Python libraries. It upscales the images to varying resolutions in order to accurately capture motion information across different scales. The tracked features are identified using the Shi-Tomasi feature detection algorithm (Shi and Tomasi, 1994). The resulting CMV field is sparse because the number of identified features is lower than the number of grid points in the map. The missing pixels are filled using a first-order interpolation method to obtain a dense CMV field.

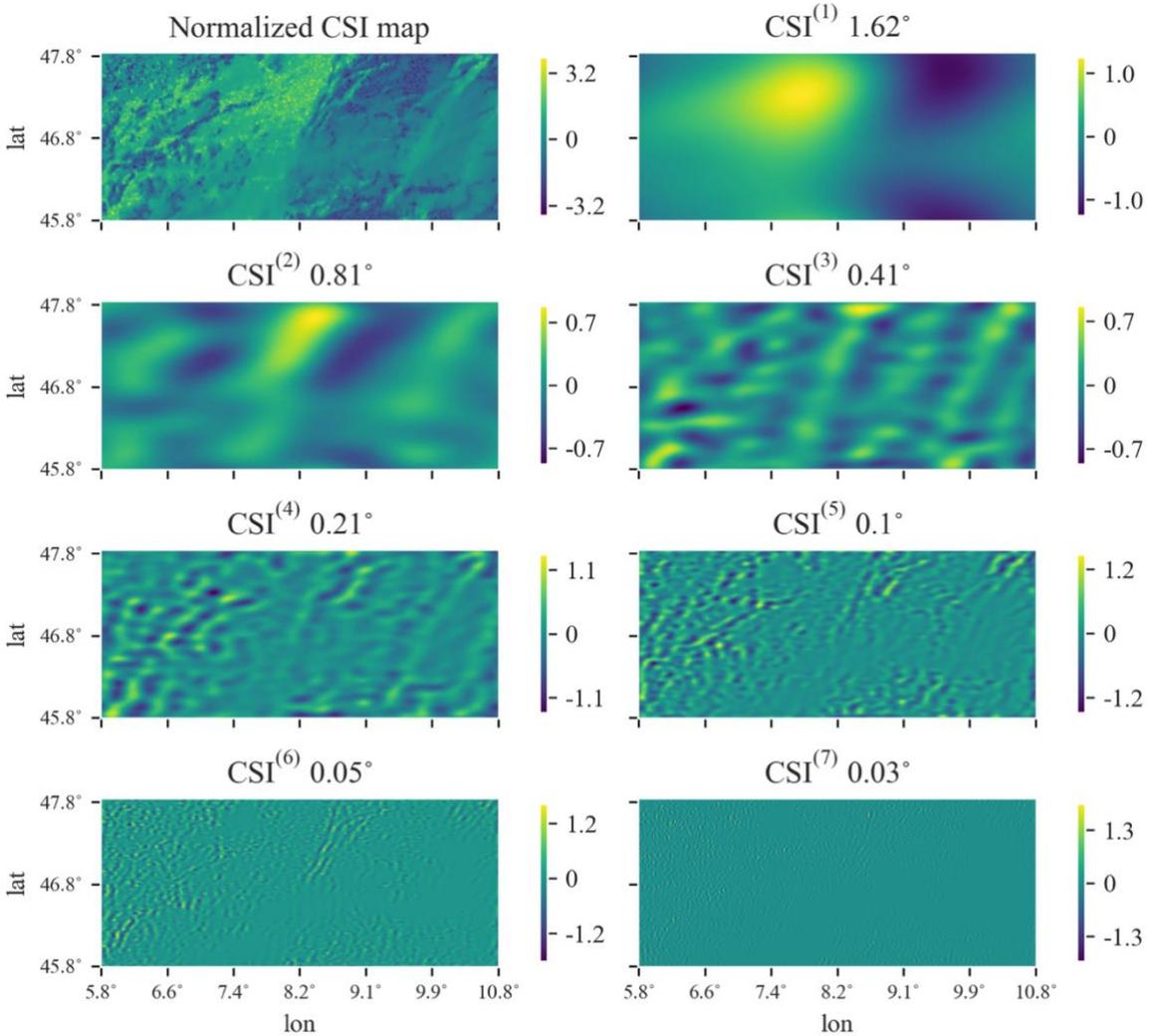

**Figure 2.** The normalized CSI map is decomposed into cascade levels in order to run a cascade-level specific AR model to accommodate scale dependent cloud life times in SolarSTEPS. Shown are a clear-sky index map (upper left panel) and seven cascade level maps (upper right to lower right panel) computed from the clear-sky

index map using a Fast Fourier transform and the seven Gaussian bandpass filters shown in Figure A5. The CSI map is derived from the SEVIRI scan of 2018-01-01 at 07:45 UTC. Fig. A5 in the Appendix shows the Gaussian bandpass filters used to decompose the CSI field.

| Cascade level c | $CSI^c$ map | Grid cell size (°) resolved by the cascade level | Spatial scale (km) resolved by the cascade level |
|---|---|---|---|
| 1 | $CSI^{(1)}$ | 1.62 | 180.2 |
| 2 | $CSI^{(2)}$ | 0.81 | 93.1 |
| 3 | $CSI^{(3)}$ | 0.41 | 46.5 |
| 4 | $CSI^{(4)}$ | 0.21 | 23.3 |
| 5 | $CSI^{(5)}$ | 0.1 | 11.1 |
| 6 | $CSI^{(6)}$ | 0.05 | 5.6 |
| 7 | $CSI^{(7)}$ | 0.03 | 3.3 |

**Table 1.** Cascade levels shown in Figure 2 for the exemplary case of seven cascade levels, and the associated spatial resolutions which correspond to the peaks of the Gaussian filters provided in Figure A5. The spatial resolution of the CSI maps is 0.02°.

## 2.3 Cloud scale-dependent autoregressive evolution

Our ensemble forecast approach goes beyond prior works in the field of probabilistic satellite-based solar forecasting in that it models the temporal evolution of the cloud radiative effect in addition to the advective motion. Our forecast approach is novel also in that it represents the temporal variability of cloudiness on different spatial scales (from ~0.02° to ~2°), outperforming state-of-the-art probabilistic advection-based solar forecasts, i.e., where probabilistic forecasts are produced via perturbation of the CMV field.

We account for the cloudiness evolution and its dependence on the spatial scale by modeling the spatial and temporal autocorrelation of the cloudiness field. Optical flow extrapolates the clouds in satellite images but it does not account for the growth and decay of cloudiness, even though these processes constitute a major source of uncertainty in SSR forecasting. Clouds follow particular scale distributions which result in associated spatio-temporal distributions of cloud shadows and sunny regions on the surface (Mol et al., 2023; Lovejoy et al., 2006). Following the work of Seed, 2003, we propose to forecast cloudiness fields by first mapping them in Lagrangian coordinates and then decomposing them into cascade levels, each of which represents a distinct spatial scale. Then, each cascade level is modeled by a separate autoregressive (AR) forecast model, as shown in step 4 of Fig. 1. The decomposition is performed by applying a Fast Fourier transform to the satellite-derived clear-sky index maps that quantify the local cloudiness and its evolution, and then applying Gaussian bandpass filters to the transformed maps. Subsequently, the Fourier-transformed and filtered maps are inverse-transformed from the frequency domain to the spatial domain. This results in cascade-level maps that correspond to different spatial scales. Fig. 2 illustrates an example of a cloudiness map and its decomposition into five cascade-level maps in the spatial domain.

Our solar forecast model provides different autocorrelation coefficients for the different spatial scales to better capture the relation between the spatial scale and temporal evolution of cloudiness. To model only the temporal evolution of cloudiness without accounting for advection, the input maps are mapped in Lagrangian coordinates by advecting the pixels along the associated cloud motion vectors. Considering an AR model of order $p$, this is done by advecting the input clear-sky index map $KI_{t0-k}$ by $(k\Delta_x, k\Delta_y) \;\forall\; k \in (1, \ldots, p)$ where $(\Delta_x, \Delta_y)$ specifies the advection (or CMV) field. The autoregressive behavior of cloudiness over time at cascade level $c \in \{1, \ldots, n_c\}$ is modeled by:

$$CSI_t^c(x,y) = \sum_{k=1}^{p} \phi_{c,k} CSI_{t-k\Delta t}^c(x,y) + \phi_{c,0}\varepsilon_t^c(x,y) \quad \forall\; c = 1, \ldots, n_c \qquad (2)$$

according to step 4 of the forecast process (Fig. 1) and similar to Bowler et al., 2006. The coefficients $\phi_{c,k}$ of the AR(p) model at cascade level $c$ are estimated using the Yule-Walker equations (Brockwell et al., 2016) as in Pulkkinen et al., 2019. The cascade levels forecasts are then merged together via summation to produce the complete forecast.

## 2.4 Noise generation

The CSI temporal evolution is modeled by the AR model defined in Eq. (2). Every cascade level $c$ is modeled by a differently parametrized autoregressive process, which is iterated to predict $CSI_t^c$. The predictions $\{CSI_t^{(1)}, ..., CSI_t^{(n_c)}\}$ are then summed to obtain the final forecast $CSI_t$. To generate a range of possible forecasts, the methodology incorporates a stochastic component in the form of a noise field $\varepsilon_t$ that is sampled and decomposed into cascade levels $\{\varepsilon_t^1, ..., \varepsilon_t^{n_c}\}$. The decomposed noise is then blended with the input CSI maps $(CSI_{t-k\Delta t}^c)_{k=1}^p$ by the corresponding AR models of the $c$ levels. To ensure realistic forecasts, the noise field must have a spatial structure similar to the CSI fields. We make use of two methods to sample spatially correlated noise: the Short-Space Fourier Transform (SSFT) method by Nerini et al., 2017 and a variation called local-SSFT, which takes also into account the spatial distribution of temporal variability in the CSI fields. Both methods employ a power-law filtering approach (Schertzer and Lovejoy, 1987) to capture the spatial distribution of cloudiness, which is described by the power spectrum of the fields. Considering a clear-sky index map $CSI$, the overall approach consists of four main steps:

1. Sampling a white noise field $\varepsilon_w \sim N(0, \Sigma)$, where $\Sigma$ is a diagonal covariance matrix,
2. Computing the Fourier transform of $\varepsilon_w$, defined by $\mathcal{F}(\varepsilon_w)$,
3. Multiplying point-wise $\mathcal{F}(\varepsilon_w)$ with the absolute value of the Fourier transformed clear-sky index field, $|\mathcal{F}(CSI)|$.
4. Then, by performing an inverse Fourier transformation, we end up with a noise field $\varepsilon$ having a spatial structure similar to $CSI$. This is because the power spectrum of $\varepsilon$ is now similar to the power spectrum of $CSI$.

The above defined approach can account for locally different spatial structures in the clear-sky index field by using local filters computed with the Short-Space Fourier Transform. In fact, SSFT computes many local Fourier transforms by using a moving window, returning many local filters to be applied to $\mathcal{F}(\varepsilon_w)$. The authors in Nerini et al., 2017 made use of a moving Hanning window to produce the local filters. They also adopted a unit variance diagonal covariance matrix to sample the white noise field ($\Sigma(i, i) = 1$). The SSFT method can reproduce the spatial structure of the CSI fields, but the intensity of the noise field is homogeneous. The local-SSFT method aims at reproducing not only the spatial structure but also the spatial distribution of temporal variability measured on the input maps $(CSI_{t-k\Delta t})_{k=1}^p$. By leveraging past information on the distribution of temporal variability, we can identify the areas in the maps which have shown enhanced variability in the past. So, the local-SSFT method samples the white noise $\varepsilon_w$ from a normal distribution $N(0, \Sigma)$, where $\Sigma_{i,i} = \sigma_i^2$ is the standard deviation of a window centered on the $i$-th pixel $(x_i, y_i)$ of the last input CSI maps. To measure $\sigma_i^2$, we compute the standard deviation in the time dimension of a moving window centered in $(x_i, y_i)$. Fig. 3 shows an example of two noise fields generated with the SSFT and the local-SSFT methods, where the same seed is used for the sampling. The local-SSFT method generates noise fields that have absolute values correlated to the temporal variability. In SolarSTEPS, the CSI map used to generate the noise field is the last input map, namely $CSI_{t_0}$.

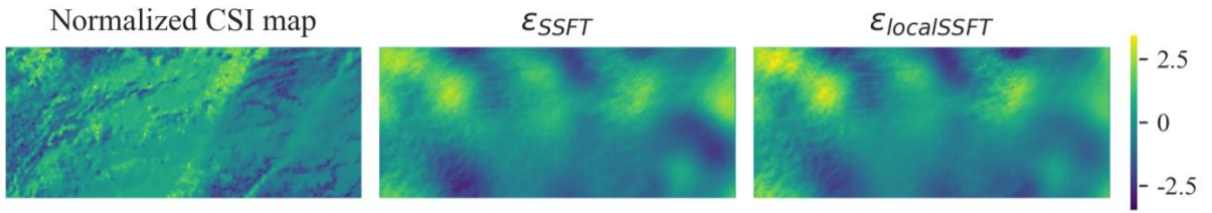

**Figure 3.** A map of the normalized clear-sky index is shown on the left. The middle panel shows the correlated noise generated by the SSFT method of Nerini et al., 2017. The right panel shows the noise generated by our local-SSFT approach. The variance of the last noise map is not homogeneous as the SSFT noise. In fact, the covariance is modeled to reproduce the variance of the input map.

## 3. Forecast model performance

The SSR forecast model presented in this study is adapted and validated to forecast clear-sky index, as discussed in section 2, which can be further easily used to predict surface solar radiation. All the models presented in our research are evaluated on the CSI forecasts, without converting to SSR. An overview of the satellite data, forecast skill scores, and benchmark models used for optimizing the forecast model parameters and for evaluating the model performance is given in the following.

### 3.1 Solar resource data

The forecasting approach is demonstrated and assessed based on satellite-derived SSR estimates over Switzerland in central Europe. The HelioMont SSR dataset (Castelli et al., 2014) is used to adapt and evaluate the forecast model. The HelioMont SSR was derived from the Spinning Enhanced Visible and InfraRed Imager (SEVIRI) on the Meteosat Second Generation satellite positioned in geostationary orbit at 0° (Schmetz et al., 2002). We refer to Carpentieri et al., 2023 for a comprehensive discussion of the accuracy and biases of the HelioMont SSR. The SSR maps have a spatial resolution of 0.02°x0.02° and a temporal resolution of 15 minutes. The SSR is derived from instantaneous intensity measurements of the SEVIRI imager which continuously measures the intensities coming from Earth in twelve visible and infrared channels, covering a field of view of ±65° N/S and ±65° W/E.

HelioMont converts the SEVIRI Level 1.5 data into SSR estimates at the respective location. It computes the SSR under clear-sky conditions to this end by using the radiative transfer model libRadtran that simulates the radiative effects of water vapor, ozone and aerosols (Mayer et al., 2005). The ERA Interim dataset (Dee et al., 2011) provides the water vapor and ozone reanalysis data used in the clear-sky SSR calculation. The aerosol optical properties originate from the Copernicus Atmosphere Monitoring Service reanalysis (Inness et al., 2019). Both datasets are provided at resolutions of 0.5° and 6 hours. Then, the all-sky SSR is computed by correcting the thus derived clear-sky SSR by the radiative effect of clouds estimated from the SEVIRI intensity measurements. HelioMont is a successor of previous Heliosat algorithms (Möser et al., 1984; Cano et al., 1986) and was developed to improve the distinction between cloud and surface albedos, especially snow cover. Cloud albedos tend to change more rapidly over time than surface albedos. HelioMont takes advantage of the difference in the typical rates of change of cloud and surface albedos for improving the distinction between clouds and snow cover. It accomplishes this by calculating the clear-sky albedo of any given day from the surface albedo and cloud mask of the past days (Stöckli, 2013). HelioMont SSR data from the year 2018 are used in this study covering 45.73˚ - 47.9˚ latitude and 5.73˚ – 10.77˚ longitude with a 0.02˚ spatial resolution, resulting in maps of size 103x241. The time resolution is 15-minute, resulting in 35040 maps. We filter out nighttime maps by removing solar zenith angles over 88 degrees and corrupted maps having more than 10% missing values. The remaining maps are then 16310. 18 images still present missing values, which are replaced by using a window-based interpolation with the nearest non-missing pixels.

$SSR$ is the product of the CSI and clear-sky SSR ($SSR_{cs}$), $SSR = CSI \cdot SSR_{cs}$. We focus on forecasting the CSI as it represents the influence of stochastic atmospheric factors on solar radiation. Moreover, the clear-sky index values do not suffer from daily and yearly seasonalities as the $SSR$, getting the CSI more suitable as forecasting variable. Moreover, the evaluation is not biased towards some periods of the year as it would be with the surface solar radiation as predicted variable due to its seasonalities.

For the test set, 72 days are selected to represent the whole year. 6 days per month are chosen, from the 5[th] to the 10[th]. From the remaining data, 500 map sequences are randomly sampled to tune the models. A sequence is intended as the combination of input maps (1 hour of data, 4 images) and output maps (4 hours of data, 16 images). The output maps are used to evaluate the quality of the model's prediction. Given the bimodal distribution of CSI maps (Fig. A6), the input images are transformed using the normal quantile transformation proposed by Bogner et al., 2012. Since the CSI distribution presents some discontinuities due to boundary conditions, a uniform noise $U(-10^{-5}, 10^{-5})$ is applied on the affected pixels in the input maps. Finally, a post-processing step is applied to the forecasted maps to match their distribution with the input maps distribution (Foresti et al., 2016).

### 3.2 Skill scores

To evaluate the forecast ensembles, we make use of deterministic and probabilistic metrics. Deterministic metrics provide a measure of bias of the ensemble mean and the quality of single ensemble members in the context of this study. Probabilistic metrics evaluate the reliability and sharpness of the forecasted probabilities (Gneiting et al., 2007, 2014; Pedro et al., 2015). A probabilistic forecast is considered reliable if the observed value falls inside the prediction interval (PI) which is given by a lower and upper bound taken from the ensemble members' values. The prediction interval coverage probability (PICP) is used to measure the reliability of our forecasts,

$$PICP = \frac{1}{N}\sum_{i=1}^{N} c_i$$

where

$$c_i = \begin{cases} 1 & \text{if } y_i \in I_i^\alpha \\ 0 & \text{otherwise} \end{cases}$$

and where $y_i$ is the actual CSI value at the pixel $I$, and where $I_i^\alpha$ is the PI with confidence level $\alpha$. $I_i^\alpha$ is defined as the interval comprised between the $\alpha$-quantile ($U_i^\alpha$) and (1-$\alpha$)-quantile ($L_i^\alpha$) of the ensemble prediction, and where N is the number of points (i.e., pixels) in a CSI map.

The sharpness measures how informative a forecast is. A forecast is sharp if it produces narrow prediction intervals. We aim to maximize the sharpness of reliable probabilistic forecasts. The sharpness is measured by the spread of the ensemble. The prediction interval normalized averaged width (PINAW) is used to quantify the sharpness of the forecasts. By setting a confidence $\alpha$, we can define the PINAW metric as the average width of the interval $I_i^\alpha$,

$$\text{PINAW} = \frac{1}{N*R}\sum_{i=1}^{N}(U_i^\alpha - L_i^\alpha)$$

where $R$ is the observations range, i.e. the maximum minus minimum CSI value.

Finally, the continuous ranked probability score (CRPS) is used to evaluate the overall quality of probabilistic SSR forecasts (Gneiting et al., 2014; Lauret et al., 2019), as it accounts for both reliability and sharpness. It does so by measuring the distance between the cumulative density function of the ensemble $F$ and the Heaviside function $I$ centered on the observation $y$. CRPS for a pixel is then defined as

$$\text{CRPS} = \int_{-\infty}^{\infty}\bigl(F(y) - I(y - \hat{y})\bigr)^2 d\hat{y}$$

Then, the CRPS is computed for all the pixels in the map grid and averaged for one step of the forecast. We consider the normalized CRPS (nCRPS) by normalizing the CRPS with the maximum clear-sky index value which is $CSI_{max} = 1.2$. Moreover, the normalized root mean square error (nRMSE) is used to measure the accuracy of the ensemble mean. To account for possible biases in the ensemble mean, the normalized mean error (nME) is considered. The nRMSE and nME for a forecasted map are defined as

$$nRMSE = \frac{1}{CSI_{max}}\sqrt{\frac{1}{N}\sum_{j=1}^{N}(\hat{y}_j - y_j)^2}$$

$$nME = \frac{1}{N * CSI_{max}} \sum_{j=1}^{N} (\hat{y}_j - y_j)$$

### 3.3 Benchmark models

We make use of three benchmark models to assess the performance of our proposed model. The first benchmark model is a simplified version of SolarSTEPS, in which the scale decomposition is not performed (SSND). The second one is a probabilistic advection (PA) model which does not account for the temporal evolution of clouds and which generates the forecast ensemble by perturbing the cloud motion field, similar to Carriere et al., 2021. The third benchmark is a deterministic persistence model (Pe).

Two baseline forecasts are employed to benchmark our forecasting approach: the Persistence model, and the Probabilistic Advection (PA) model which both lack a representation of cloudiness processes. The Persistence forecast model predicts the same value observed at the forecast time $t_0$,

$$CSI_{t_0+h} = CSI_{t_0} \quad \forall \ h = 1, \cdots, n.$$

The PA model generates the ensemble of forecasts of clear-sky index maps using a Monte Carlo approach. The PA model does not account for the temporal variability of the clear-sky index, and instead of perturbing the temporal evolution, the PA model applies noise to the motion field at the initial stage of the forecast. A noise field is sampled and added to the LK motion field $(v_x, v_y)$ to generate an ensemble member. The perturbed motion field is then applied to advect the clear-sky index (i.e., cloudiness) maps. This step is repeated $k_{ens}$ times to generate the ensemble of forecasts. For comparability, the noise is sampled following a similar approach as the one used in Carriere et al., 2021. Two perturbation fields are sampled from a Gaussian and a von Mises (vM) distributions, $\varepsilon_r \sim N(0, \sigma_r^2)$ and $\varepsilon_\theta \sim vM(0, k)$, and applied to the magnitude and angle of the motion vectors, respectively, generating a perturbed motion field. The perturbed motion fields obtained in this way are then used to advect the clear-sky index maps, creating an ensemble of different forecasts. To maintain the positivity of the norm of the motion vectors, the noise is iteratively sampled until the perturbed norm field has only positive values or a maximum number of iterations is reached. Eventual negative values at the end are mapped back to zero. Moreover, to highlight the effectiveness of our scale-dependent approach, a simplified version of SolarSTEPS (SSND) is used as a comparison, in which no decomposition is performed. Table 2 provides a summary of the probabilistic forecast models compared in this study.

| Model | Abbreviation | Description |
|---|---|---|
| SolarSTEPS | SolarSTEPS | Probabilistic advection-based forecasting model with temporal cloud evolution simulation. The forecast ensemble is generated by perturbing the scale-dependent autoregressive models with spatially correlated noise. |
| Probabilistic advection | PA | Probabilistic advection-based forecasting model. The forecast ensemble is generated by perturbing the cloud motion field norm and direction. The perturbed motion fields are then used to advect the last input map in different ways. |
| SolarSTEPS without decomposition | SSND | Probabilistic advection-based forecasting model with temporal cloud evolution simulation. The forecast ensemble is generated by perturbing the autoregressive model with spatially correlated noise. No decomposition is performed so only one AR model is used to simulate the temporal evolution of cloudiness. |
| Persistence | Pe | Deterministic forecasting model. It works by propagating the last seen map into the future. |

**Table 2.** Summary of the SSR forecast models: SolarSTEPS and three benchmark models.

## 4. Results and discussion

Our probabilistic solar forecast model SolarSTEPS is tested with SSFT and local-SSFT ensemble generation and compared to the benchmark models (Table 3). The test set is composed of six consecutive days selected from each month, specifically, the 5th to the 10th day of each month in 2018. These time periods are randomly selected when defining the test set. Six consecutive days are chosen from each month to ensure that the test set is representative of all seasons and weather conditions. All daytime periods from sunrise until four hours before sunset are used as input to initiate the forecasts, with sunrise (sunset) being defined by the time when the solar zenith angle falls below (exceeds) 88°. The resulting test dataset consists of 1818 sequences, each one with 20 associated Meteosat scans and derived cloudiness maps (4 for the input and 16 to evaluate the output), spanning 72 days in total. We do not run forecasts starting in the last four hours of each daytime period because they would not be fully validatable due to the upcoming sunset and the resulting lack of satellite-derived CSI to use as ground truth.

Optimal model parameters are determined for all compared forecast models by first randomly sampling 500 sequences from the daytime periods of the remaining days of 2018, again considering only times that are more than four hours forecasting before sunset, and then running probabilistic on these 500 instances for different model parameter combinations. The optimization procedure and the optimized model parameters are summarized in Section A1 of the appendix and Table 3, respectively. The model parameter optimization is performed before and independently of the final evaluation of the models on the test dataset. This section discusses the performances of the models on the test dataset. The number of ensemble members is set to 40 for the model performance assessment on the test set because 40 ensemble members correspond to a reasonably sized ensemble and at the same time a low continuous ranked probability score (CRPS) on the validation set, as shown in Fig. A7.

| Model | Abbreviation | Optimal parameters |
|---|---|---|
| SolarSTEPS with SSFT ensemble generation | SolarSTEPS$_{SSFT}$ | • AR order: 1<br>• Number of cascade levels: 9<br>• Noise generation method: SSFT<br>• SSFT window size: 35 |
| SolarSTEPS with local-SSFT ensemble generation | SolarSTEPS$_{local-SSFT}$ | • AR order: 1<br>• Number of cascade levels: 7<br>• Noise generation method: local-SSFT<br>• SSFT window size: 13<br>• Local window size: 47 |
| SolarSTEPS no decomposition | SSND | • AR order: 1<br>• Number of cascade levels: none<br>• Noise generation method: SSFT<br>• SSFT window size: 57 |
| Probabilistic advection | PA | • $\sigma_r^2$: 9.23<br>• $k$: 0.15 |

**Table 3.** The parameter values resulting from the model parameter optimization process described in section A1. SSFT window size refers to the size of the Hanning moving window used in Nerini et al., 2017 for the SSFT ensemble member generation method. Local window size refers to the size of the moving window used to compute the covariance diagonal matrix defining the multivariate Gaussian distribution from which we sample the white noise. For the PA model, $\sigma_r^2$ is the variance of the noise field used to perturb the norm of the CMV field whereas $k$ is the parameter governing the Von Mises distribution from which we sample the noise applied to perturb the direction of the CMV field.

We show that the temporal evolution of cloudiness (intended as cloud radiative effect) depends on its spatial extent (Fig. 4). The temporal variability of cloudiness, as measured by the clear-sky index, depends on the considered spatial scale. Cloudiness on smaller spatial scales shows higher temporal variability than cloudiness on larger spatial scales, as illustrated in Fig. 4. To demonstrate this, we applied a similar approach as Venugopal et al., 1999, who compared two consecutive precipitation maps at different spatial scales through downscaling. We quantified the temporal variability at different spatial scales by first decomposing the clear-sky index maps with FFT and Gaussian bandpass filters into cascade-level maps (step 3 of the SolarSTEPS forecast process, as shown in Fig. 2) and then computing the squared Pearson correlation coefficient of each cascade level for two successive time steps $t$ and $t + 1$, as shown in Fig. 4. Our procedure is the same as computing the lag-1 autocorrelation in time for all the cascade levels. The procedure is computed in Lagrangian coordinates to exclude advection as a potential source of variability. The validation set (section 3) is used to estimate the distribution of the autocorrelation that quantifies the temporal variability of cloudiness at the different spatial scales (Fig. 4). The result highlights that cloudiness is less variable in time at larger spatial scales, showing autocorrelation values close to 1, as the pixel brightness changes more slowly in time than at smaller spatial scales where it reaches values close to 0.4. These observations provide motivation to us for applying distinct autoregressive (AR) models for different spatial scales to capture the scale-dependent temporal autocorrelation in the clear-sky index fields, i.e. in the cloudiness. Fig. A8 provides the same plot for different numbers of cascade levels, confirming our results.

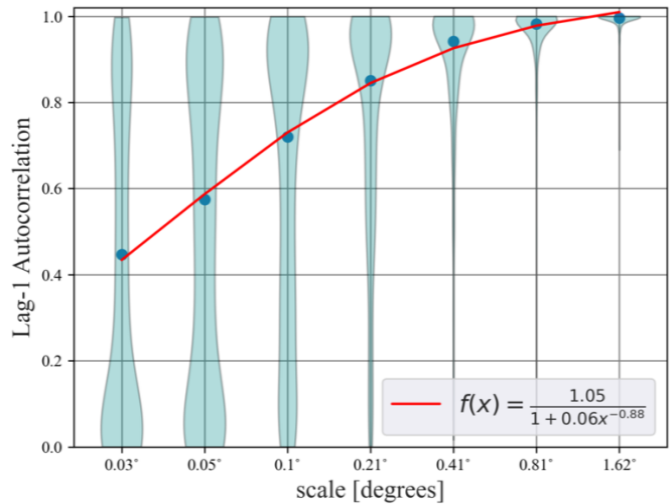

**Figure 4.** Kernel density estimates (violin plots) of the autocorrelation distribution (Pearson correlation coefficient) derived from comparing cascade-level maps at two successive time steps based on 500 randomly sampled pairs of clear-sky index maps. The horizontal axis states the spatial scales of the five cascade levels for which the lag-1 autocorrelation distributions were estimated. The blue dots represent the average correlation coefficient for each cascade level. To compute the lag-1 autocorrelation values, the clear-sky index maps are decomposed into the five cascade levels resulting from the Gaussian bandpass filtered FFT (step 3 in Fig. 1), each one representing a different spatial scale. Then, two consecutive maps are mapped in Lagrangian coordinates in order to account only for the temporal variability at different scales, and the correlation is defined by the Pearson coefficient. The validation set is used for the analysis. The blue line represents the average autocorrelation for each cascade level. Different numbers of cascade levels are tested, and all of them confirm the scale-dependence of the temporal variability of the CSI.

The normalized continuous ranked probability score (nCRPS) is used for quantifying the ensemble forecast quality of the compared probabilistic models (Table 3). The SolarSTEPS models (with SSFT and local SSFT ensemble generation) are compared to two probabilistic benchmark models: the probabilistic advection model (PA) which does not account for the autoregressive character of CSI, and a probabilistic advection model that accounts for the autoregression but not the scale dependence of CSI (SSND). Fig. 5 provides the nCRPS computed and averaged on the test set. The SolarSTEPS forecast quality clearly exceeds that of the benchmark models' forecasts at all lead times, showing a significantly lower nCRPS (upper panel). The performance improvement with regard to the probabilistic advection model increases monotonously with lead time. When compared to the benchmark models, SolarSTEPS extends the forecast lead time by up to 45 minutes at a given ensemble forecast quality, as can be seen by comparing SolarSTEPS and the benchmark models at constant nCRPS in the upper panel of Fig. 5. It is also apparent that the local SSFT ensemble generation slightly improves the performance of SolarSTEPS compared to the SSFT based ensemble generation.

The distribution of the nCRPS computed on the test set is shown in the lower panel of Fig. 5. Each distribution is composed of 1818 nCRPS values in accordance with the test set size. SolarSTEPS not only provides a lower mean nCRPS but its entire nCRPS distribution is shifted to lower values when compared to the PA and SSND benchmark models. SolarSTEPS also has a narrower nCRPS distribution than the benchmark models, so its forecast skill varies less among the different instances of the test set. In contrast, PA ensemble forecasts result in the largest errors on the test sets.

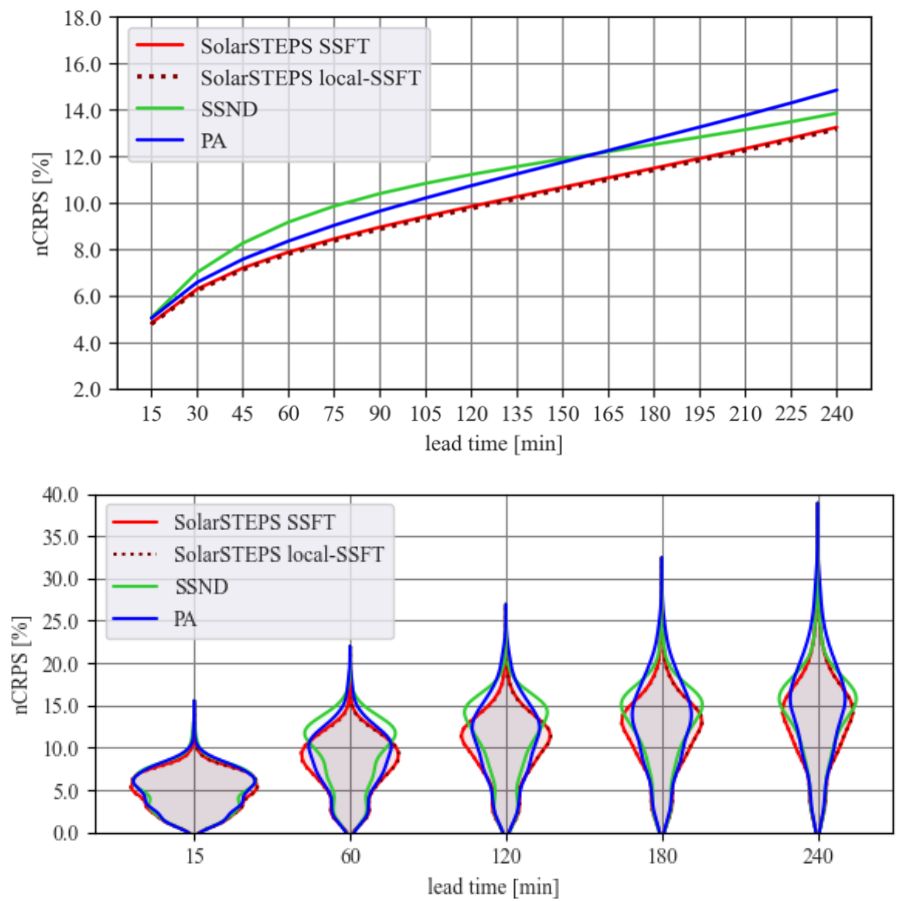

**Figure 5. Upper panel** shows the normalized CRPS computed on the test set, with the CRPS values averaged over all CSI map pixels of every forecast performed on the test set. SolarSTEPS outperforms the baseline models (Probabilistic advection and SSND) at all the lead times. The performance improvement with regard to probabilistic advection increases with the lead time. **Lower panel** shows the distributions of the nCRPS values across the test set for each of the forecast models at different lead times. The nCRPS distributions were derived by a kernel density estimation. Each distribution is composed of 1818 nCRPS values in accordance with the test set size.

The spatial distribution of nCRPS is shown in Fig. 6. SolarSTEPS outperforms the PA and SSND benchmark models, showing lower nCRPS especially for longer lead times. The nCRPS is correlated with altitude in all four models, so the forecast skill tends to be reduced in the mountain regions of Switzerland, and higher in valleys and along the Swiss plateau. This may partly be due to the fact that optical flow models do not recognize mountain barriers. Biases in the satellite-derived SSR estimates and the derived CSI may also play a role (Carpentieri et al., 2023). The Probabilistic Advection and SSND models exhibit higher CRPS than SolarSTEPS at all considered forecast lead times, as shown in Figs. 5 and 6. Modeling the cloud evolution with scale-dependent autoregressive processes significantly improves the forecast skill at all lead times. SolarSTEPS' advantageous forecast performance becomes apparent progressively as forecast lead times increases when compared to the Probabilistic Advection (PA) and SSND models.

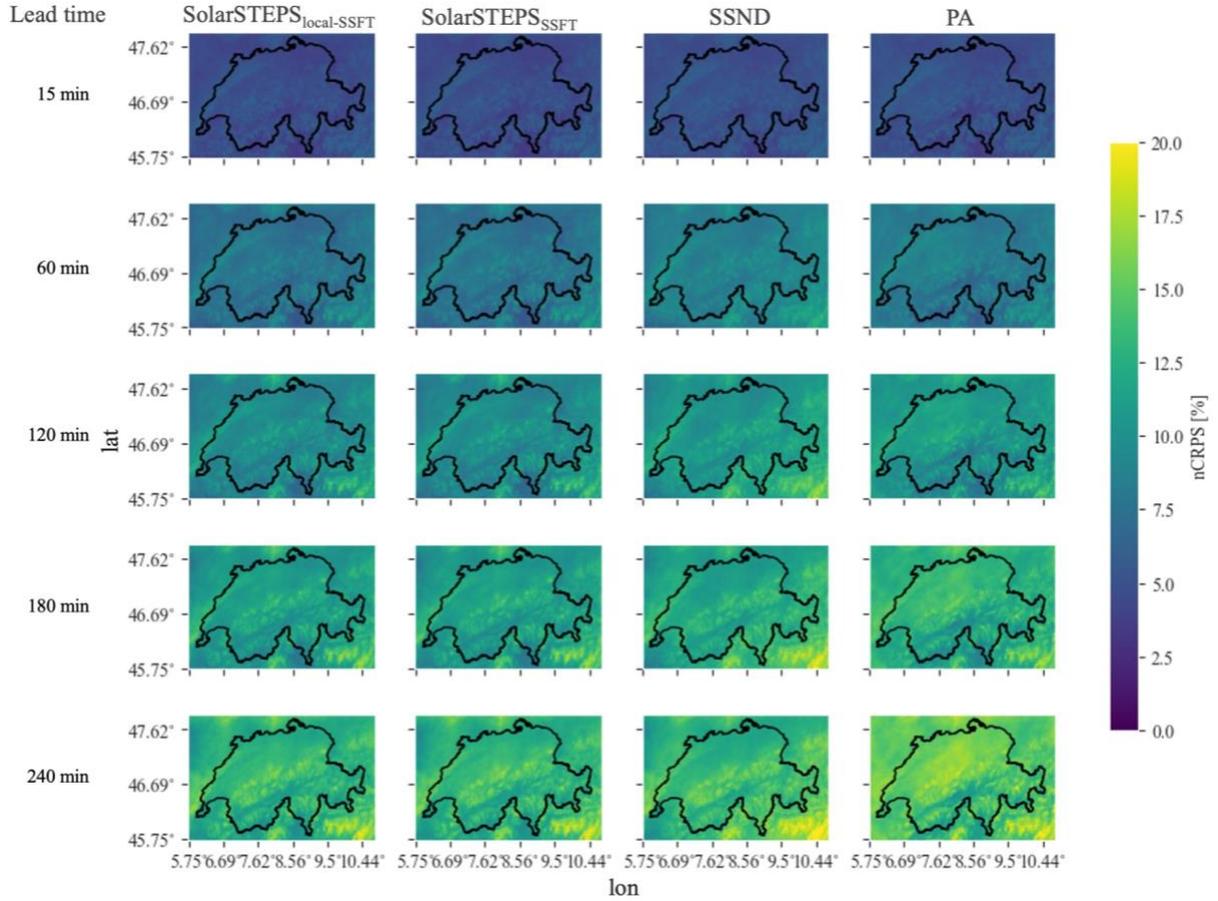

**Figure 6.** Pixelwise average normalized CRPS computed on the test set. The black contour line shows Switzerland. Modeling the cloudiness evolution significantly improves the quality of probabilistic solar forecasts.

The scale-dependent forecasts of SolarSTEPS significantly improve the forecast skill in all weather conditions, as highlighted in Fig. 7. We define cloudy (or sunny) conditions based on the first (or last) quartile of the test set instances ordered by their average CSI value. To define the sunny and cloudy test set instances, the average CSI value is computed for each input CSI map, yielding 1818 average CSI values. Then, those values and associated CSI maps are ordered and the first and last quartiles are kept as sunny and cloudy instances, respectively. All four models perform worse in cloudy weather as cloudy conditions tend to be associated with a more variable CSI and are harder to forecast than sunny time periods. The forecast performance improvement of SolarSTEPS is particularly significant in cloudy conditions. SolarSTEPS reduces the nCRPS by several percent in absolute terms compared to the state-of-the-art PA model and even the SSND model: The nCRPS improves from about 13% to 11%, and from about 16% to 14% at lead times of two and four hours, respectively, in cloudy conditions. SolarSTEPS with local-SSFT noise generation has a 14.1% and 10.2% lower nCRPS than the SSND and PA forecast ensemble model, respectively. The local-SSFT ensemble generation method accomplishes forecast improvements mainly in sunny conditions when compared to the SSFT method. This is probably due to the ability of the local-SSFT noise generation to focus on the most variable areas of the map. In fact, in sunny situations the clouds are sparse and they are the main source of uncertainty. Local-SSFT can identify these maps and generate a noise field with higher variability in these areas. We also highlight the better performance of the local-SSFT noise generation in a low-variability and sunny situation with Fig. A9, where all 4 models are run to forecast four hours of CSI maps (from 2018-07-09 07:30 am to 2018-07-09 11:30 am UTC+0). In the example, local-SSFT can better identify the most variable areas in the map than SSFT, especially the clouds in the south and northeast part of the map. We find similar results when looking at the variability of the input maps as measured by the standard deviation. SolarSTEPS outperforms the benchmark models by 8% and 15% (PA and SSND) in highly variable weather conditions (Fig. A10). The forecast performance is better in low-variability weather conditions, resulting in a lower nCRPS for all four models and a smaller performance improvement of SolarSTEPS in absolute terms. In summary, SolarSTEPS outperforms the benchmark models in cloudy and highly variable conditions, while the four models show similar behaviors in low variability and sunny weather with a moderately lower nCRPS of SolarSTEPS.

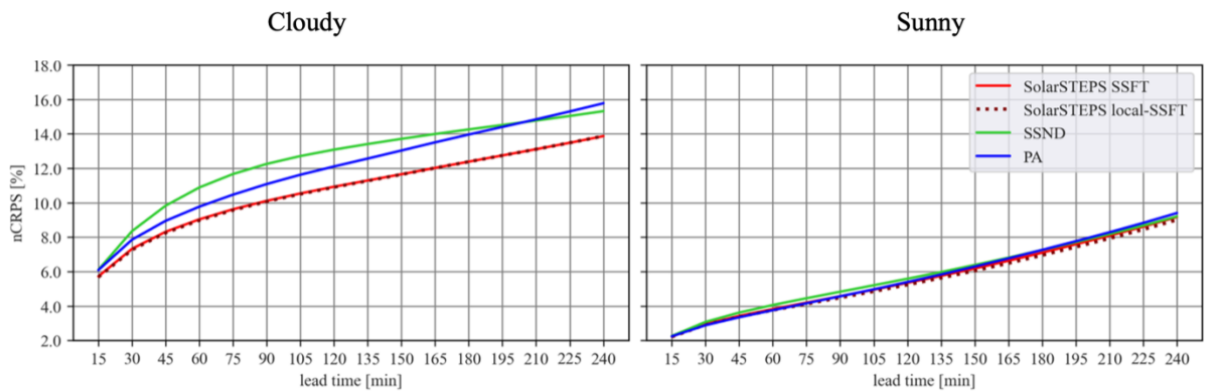

**Figure 7.** Average normalized CRPS of forecasts in different weather conditions. The nCRPS is computed as in Fig. 5 but on different subsets of the test set that correspond to cloudy or sunny weather. Cloudy (sunny) conditions are defined by CSI values in the first (last) quartile of the test set instances ordered by their average CSI value as computed on the input CSI maps.

The forecast ensemble means are evaluated in Fig. 8. The normalized root mean square errors (nRMSE) and mean errors (nME) of the forecasts are computed with regard to the ensemble mean for the probabilistic models and with regard to the deterministic persistence forecasts. The ensemble means of the SolarSTEPS forecasts on the test set show the lowest nRMSE compared to the benchmark models PA, SSND and the Persistence forecast. SolarSTEPS$_{local-SSFT}$ has the lowest absolute bias (nRMSE), providing an average 14.7%, 5.4%, and 8.7% improvement over the Persistence, Probabilistic Advection and SSND models, respectively.

From the normalized mean error plot, we can see the slight negative bias in the test set's output images from the positive bias of the Persistence model. Overall, SSND shows the lowest nME, possibly due to a compensation of negative and positive biases.

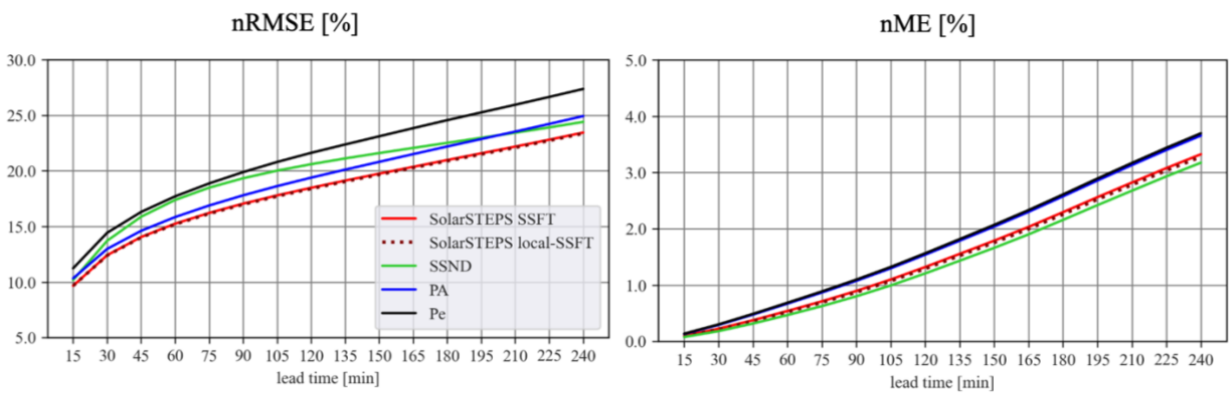

**Figure 8.** Average normalized RMSE computed on the test set. The mean of the ensemble is used for the metric computation for the probabilistic models. The three models are compared also to the Persistence model (Sec. 3.3). SolarSTEPS shows a lower nRMSE for all the lead times, with an average 13.8%, 4.5%, improvement w.r.t. Persistence and Advection model, respectively.

We evaluate the reliability and sharpness of the ensemble forecasts of SolarSTEPS and the benchmark models (Tab. 3) using the prediction interval coverage probability (PICP) and the prediction interval normalized averaged width (PINAW). A 90% confidence interval is applied ($\alpha$ is set to 0.05). Fig. 9 shows the PICP and PINAW.

Compared to SolarSTEPS, the SSND model generates ensembles with significantly wider prediction intervals, resulting in less informative probabilistic forecasts, while PA results in narrower prediction intervals but generates less reliable forecasts. The noise generation method seems to not significantly affect the prediction interval of SolarSTEPS, which results in a trade-off between SSND and PA.

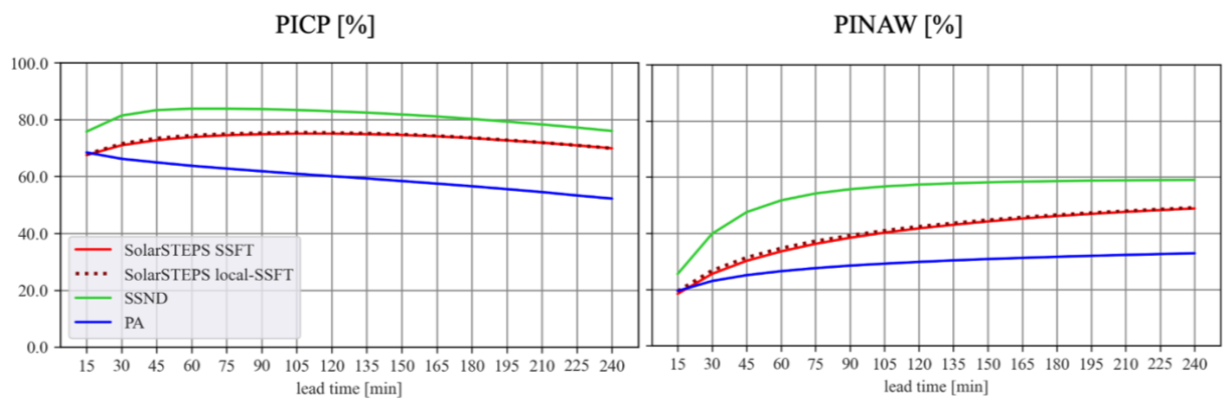

**Figure 9. Left panel** shows the average PICP computed on the test set for lead times of up to four hours. Both SolarSTEPS models outperform the PA and SSND forecasts. **Right panel** shows the average PINAW computed on the test set.

The same results are shown in the reliability diagrams in Fig. 10. Reliability diagrams plot observed probability against forecast probability. So in perfectly reliable forecasts, the observed and forecast probabilities should be equal, and so their curve would lie on the diagonal line in the reliability diagram. The diagrams are computed on the entire test set, for three thresholds (0.3, 0.6, 0.9), where the values are relative to the probabilities of exceeding these thresholds by the models. Fig. 10 shows that all forecast models yield positively sloped curves, indicating that all models provide skillful forecasts. The slope of the PA forecast curve substantially deviates from the diagonal, whereas the curves of the SolarSTEPS and SSND forecasts follow the diagonal line closely. SSND appears to be the most reliable model. On the other hand, PA shows an overconfident behavior, due to the very narrow prediction interval. Also, SolarSTEPS seems to be slightly overconfident. To conclude, SSND showed to be the most reliable model by generating wide prediction intervals, while PA computes narrow ensemble forecasts. Both models have problematic drawbacks. The first cannot provide useful probabilistic forecasts, while the second cannot reproduce the high variability present in the CSI maps. For these reasons, SolarSTEPS outperforms the benchmark models by providing the most useful probabilistic forecasts in both sunny and cloudy situations as measured by the CRPS. Moreover, we showed that by introducing temporal variability information in the noise field generation, we can increase the model performance in low-variability situations.

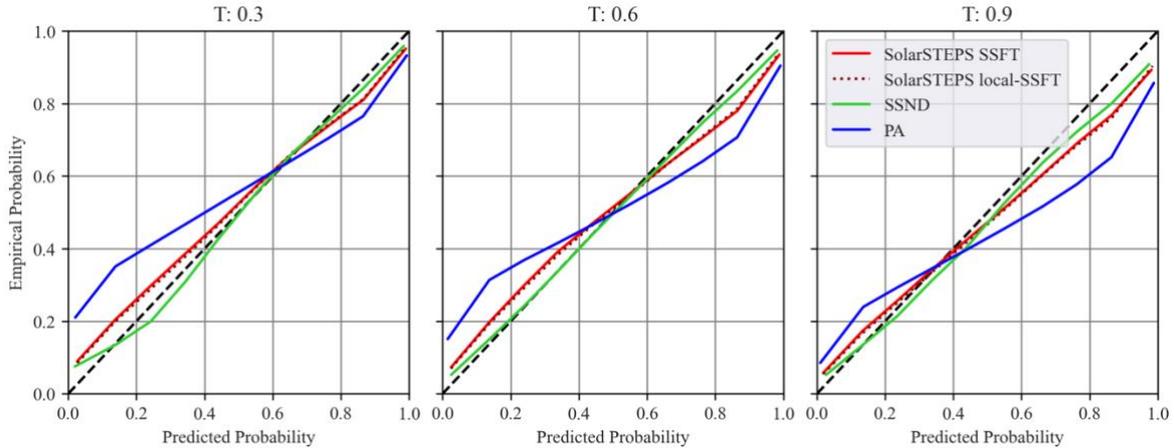

**Figure 10.** The reliability diagrams show observed probability (vertical axis) plotted against forecast probability (horizontal axis). The diagrams are shown for three different thresholds T (0.3, 0.6, 0.9). The thresholds are used to derive probabilities of exceeding them from the forecast ensemble. The probabilistic cloud evolution modeling (SolarSTEPS and SSND) strongly increases the reliability of the ensemble predictions compared to probabilistic advection (PA) model.

## 5. Conclusions

We show that the temporal evolution of the cloud radiative effect depends on the spatial scale of clouds. We introduce a probabilistic optical-flow-based forecast model which accounts for the temporal evolution and spatial structure of cloudiness in addition to advection. Our model, SolarSTEPS, is the first probabilistic satellite-based nowcasting model that accounts for the temporal evolution of cloudiness. It is also the first such model to account for the scale dependence of the cloudiness' temporal variability.

We demonstrate that modeling the cloudiness evolution – in addition to advection – significantly improves the forecast quality compared to state-of-the-art probabilistic advection models, which do not model the cloudiness evolution. SolarSTEPS yields a 9.3% improvement in ensemble forecast quality (nCRPS) compared to probabilistic advection. This corresponds to an extension of the forecast lead time by about 45 minutes at a given level of forecast quality, measured by the CRPS.

We also show that predicting the cloud evolution at different scales further improves the forecast quality, as seen by comparing SolarSTEPS to SSND forecasts. Moreover, we demonstrate that incorporating a local covariance matrix in ensemble generation further improves the forecast quality because it accounts for local variability. SolarSTEPS forecasts clearly yield improved forecast quality (lower nCRPS) when compared to probabilistic advection and SSND benchmark forecasts. The forecast quality improvements are particularly pronounced in cloudy and variable weather conditions.

In conclusion, our forecast model captures the autocorrelated nature and size-dependent evolution of cloudiness in addition to cloud advection, and thereby achieves a significant improvement in ensemble forecast quality compared to the state-of-the-art. Our SolarSTEPS model can predict cloudiness for the full area covered by a satellite image, so its forecasts are a suitable basis for predicting regionally aggregated PV production. Finally, SolarSTEPS requires only about 5 seconds to produce a forecast with 40 ensemble members at 16 forecast lead times if given CSI maps of approximately 25'000 pixels (corresponding to the MSG view of Switzerland shown in Fig. 6). Its fast forecast computation and superior forecast quality make SolarSTEPS highly suitable for operational nowcasting. Future research should investigate the utilization of different autoregressive orders for the different scales. In fact, for the bigger scales higher order AR models can be used to emulate the temporal evolution. Following similar progresses in precipitation nowcasting, integrated autoregressive models (ARI) showed better performance in deterministic forecasts (Pulkkinen et al., 2020) and could be applied in SolarSTEPS.

**Code Availability**

The code for SolarSTEPS is made available at https://github.com/albertocarpentieri/SolarSTEPS.

## Acknowledgements

We acknowledge valuable discussions with Laurent Vuilleumier. C. A. and A. M. acknowledge funding from the Swiss National Science Foundation (grant 200021_200654). The HelioMont data are licensed and can be obtained from the MeteoSwiss customer service via https://www.meteoswiss.admin.ch/home/form/customer-service.html.

implications for grid integration: Towards carbon neutrality, Renewable and Sustainable Energy Reviews, 161, doi:10.1016/j.rser.2022.112348.

## Appendix.

## Implementation details

The SolarSTEPS Python implementation is based on the Pysteps library for precipitation nowcasting (Pulkkinen et al., 2019) that we adapted for solar resources nowcasting applications. The main difference between precipitation and cloudiness nowcasting lies in the fact that in our case, the aim is to forecast the whole field, while for precipitation, the model focuses on a subset (i.e. the precipitation area). Moreover, the CSI distribution drastically differs from the log-normal distribution of precipitation requiring the adaptation of the pre-processing step. To normalize the CSI maps, we adopt a quantile transformation to map the bimodal distribution into a normal one. Moreover, we apply a uniformly sampled perturbation to the maps before the quantile transformation to achieve a continuous density distribution. The HelioMont CSI maps can, in fact, have multiple pixels with the same value due to boundary constraints in the CSI computation.

Regarding noise generation, the SSFT noise implementation is based on the Pysteps library, to which we added the possibility to generate white noise with different covariance diagonal matrices resulting in the local-SSFT noise generation. In SolarSTEPS it is possible to compute the covariance matrix with different window sizes. To manage the borders, a mirroring approach is implemented.

A significant modification is made in the autoregression step of the model. Following Eq. (6) of Pulkkinen et al., 2019 the output of the autoregression model converges to a map of unit variance if both input and noise have unit variance. We enforced the distribution stationarity by normalizing the output of the AR model at every forecast step. The extra normalization step is applied after the AR iteration of every cascade level. In fact, the input of the AR model are the cascade levels of the CSI input and the noise, which are all normalized to zero mean and unit variance. However, after the AR iteration the distribution of the output can be different, so we enforce the stationarity by applying a normalization step. The result of this modification is shown with a forecast example in Fig. A1. By enforcing the distribution stationarity for the different cascade levels, the forecast looks more realistic and obtains better performance.

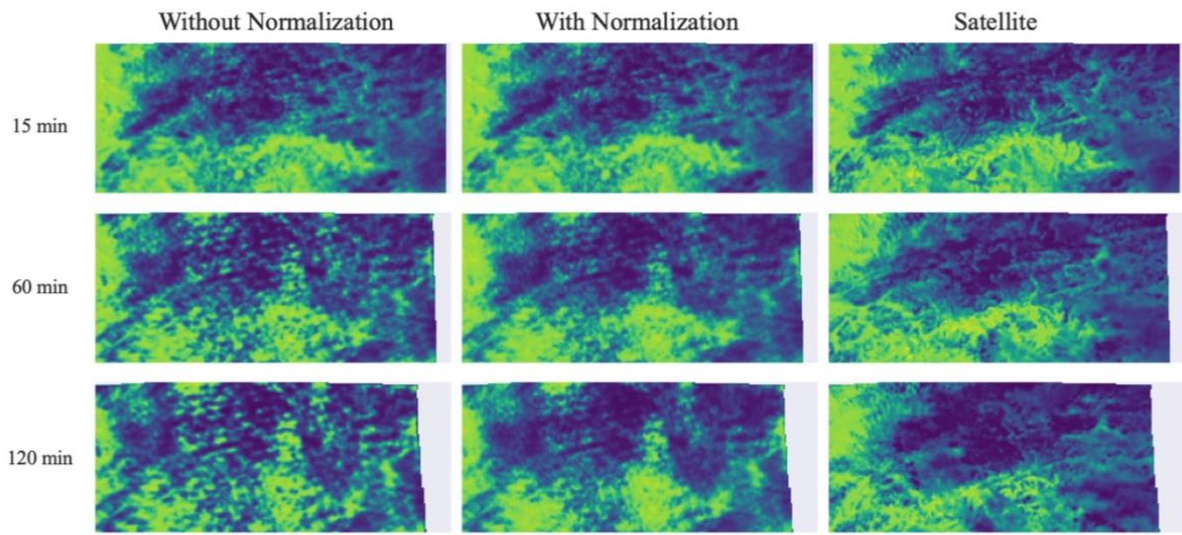

**Figure A1.** Two forecasts computed with SolarSTEPS with and without the AR normalization step, respectively. The forecasts are relative to one ensemble member. The right subpanel shows the actual maps corresponding to the forecasted lead times. The normalization step preserves the variance in the cascade level avoiding some cascade levels to increase/decrease their variance. An increase or decrease of the cascade levels' variance changes the power spectrum leading to some scales to dominate the final forecast. In the example, the image is decomposed in 9 cascade levels, with a variance increase of the $4^{th}$, $5^{th}$ and $6^{th}$ cascades leading to artificial objects in the final forecasts.

# Model parameters optimization

The three probabilistic models are tuned on 500 sequences randomly sampled with no repetition from 2018 data, i.e., the validation set. This step is performed to optimize the models before the final evaluation on the test set. The minimized metric is the average CRPS computed on all 16 output maps for all the sequences composing the validation set. The ensemble size is chosen to be 10. The algorithm used for the optimization is Adaptive TPE implemented in the Hyperopt Python library (Bergstra et al., 2013). The optimized parameters for the Probabilistic Advection model are the variances of the Gaussian and von Mises noise fields applied to the motion field's norm and angle, respectively. The search spaces are defined by the lower and upper bounds. The number of optimization steps is set to 50. The results of the optimization are shown in Tab. A1 and Fig. A2.

| Parameters | Search space | Optimal value |
|---|---|---|
| $\sigma_r^2$ | [0, 20] | 9.29 |
| $k$ | [0, 20] | 0.17 |

**Table A1.** Optimization table for the Probabilistic Advection model. The parameters tuned are relative to the variances of the Gaussian and von Mises distributions used for the generation of the noise fields for the norm and angle of the advection field.

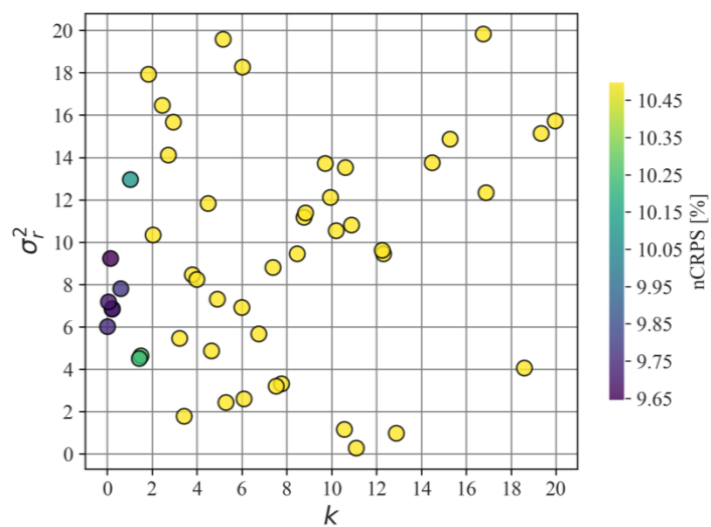

**Figure A2.** Optimization results for PA. $k$ is the parameter governing the Von Mises distribution from which we sample the noise to perturb the direction of the vectors in the motion field. $\sigma_r^2$ is the variance of the Gaussian distribution from which we sample the perturbing noise used to perturb the norm of the motion field. Every dot in the

For SolarSTEPS, the optimized parameters are shown in Tab. A3-A4 and are the autoregressive order (AR order), the noise generation method, the number of cascade levels and the size of the Hanning windows ($s_h$) used for the SSFT noise generation. In the case of local-SSFT, the window size used to compute the covariance matrix ($s_{std}$) for the white noise is also optimized. For SSND, the number of cascade levels is fixed to one and the noise generation is performed with the SSFT method. The number of evaluations is set to 50 for SolarSTEPS[local-SSFT] and 30 for SolarSTEPS[SSFT] and 20 for SSND. The optimization algorithm is Adaptive TPE from the hyperopt library (Bergstra et al., 2013). Every evaluation corresponds to the evaluation of a single parametrization of the model on the entire validation set. Table A2-A3 shows the optimization results for the two versions of SolarSTEPS and the correspondent search spaces. The search spaces for $s_h$ and $s_{std}$ are only the odd number comprised by the bounds in order to have a centered moving window. The results of all the optimization runs are shown in Fig A3-A4. For both noise generation methods, the optimal AR order is 1. Also, the cascade level $n_c$ has a significant influence on the model performance, having the optimal values ranging from 7 to 9.

| Parameters | Search space | Optimal value |
|---|---|---|
| AR order | {1, 2} | 1 |
| $s_h$ | [11, 101] | 57 |

**Table A2.** Optimization table for SSND. The parameters tuned are the order of the autoregressive model and the size of the SSFT window ($s_h$).

| Parameters | Search space | Optimal value |
|---|---|---|
| AR order | {1, 2} | 1 |
| $s_{std}$ | [11, 101] | 47 |
| $s_h$ | [11, 101] | 13 |
| Cascade levels | [4, 9] | 7 |

**Table A3.** Optimization table for SolarSTEPS$_{local\text{-}SSFT}$

| Parameters | Search space | Optimal value |
|---|---|---|
| AR order | {1, 2} | 1 |
| $s_h$ | [11, 100] | 43 |
| Cascade levels | [4, 9] | 9 |

**Table A4.** Optimization table for SolarSTEPS$_{SSFT}$.

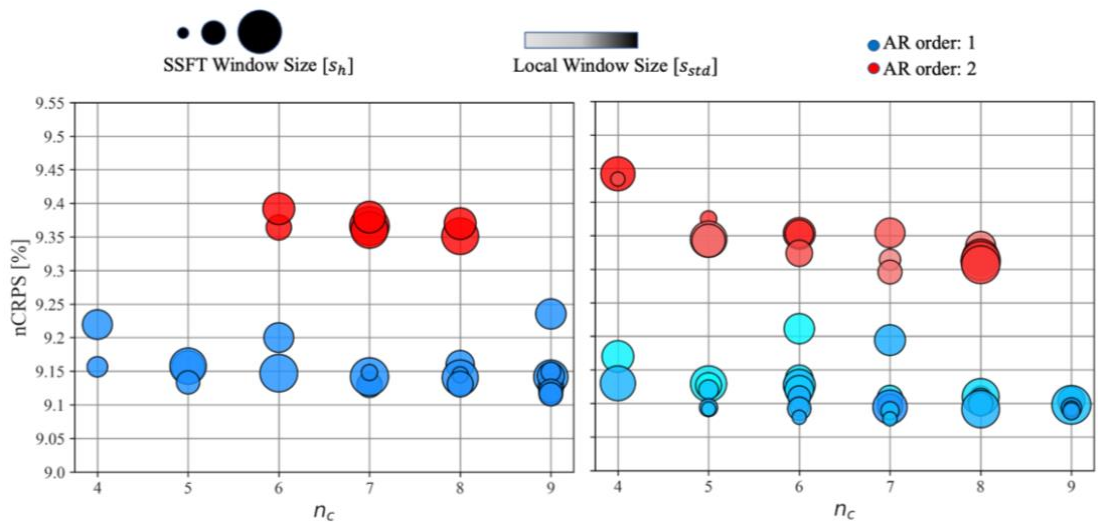

**Figure A3.** Optimization results for SolarSTEPS$_{SSFT}$ (left panel) and SolarSTEPS$_{local\text{-}SSFT}$ (right panel). The number of optimization steps is different as the two models have different numbers of parameters to optimize.

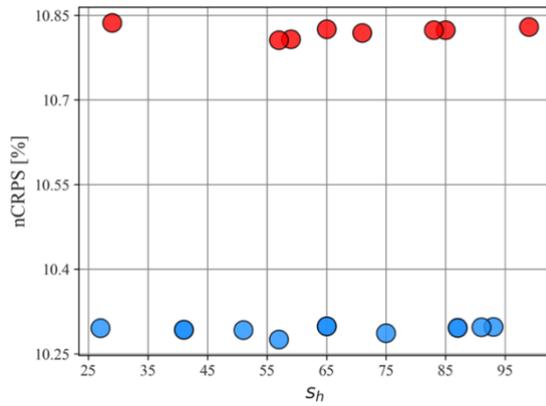

**Figure A4.** Optimization results for SSND model. As in Fig. 3 the blue dots represent the evaluations with autoregressive order equal to 1, which results in a significantly better performance compared to the second order runs (red dots).

# Additional Figures

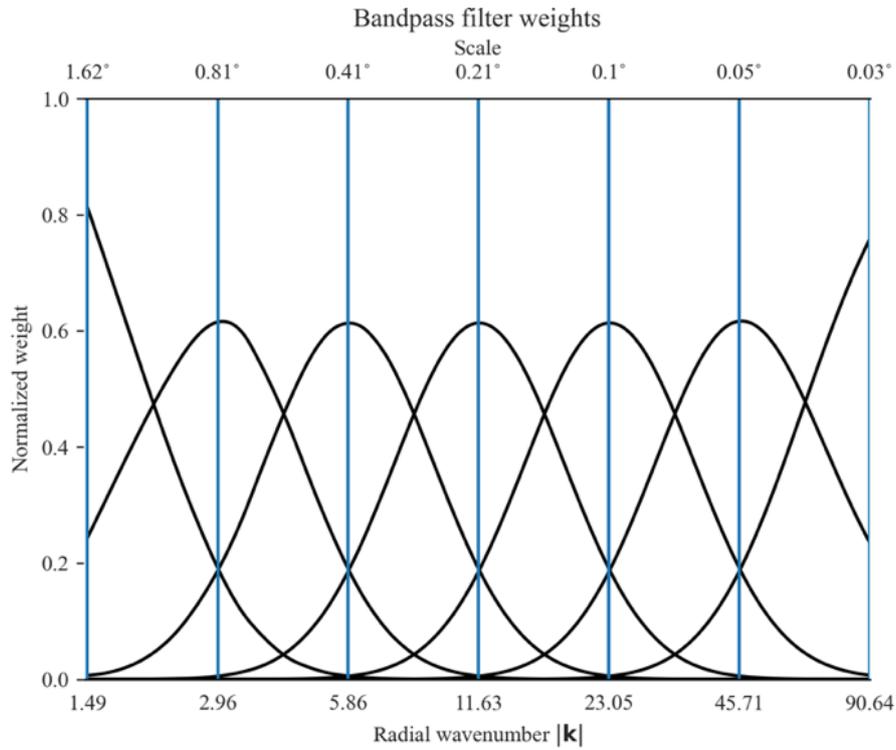

**Figure A5.** Gaussian bandpass filters used to decompose a CSI map into seven cascade levels.

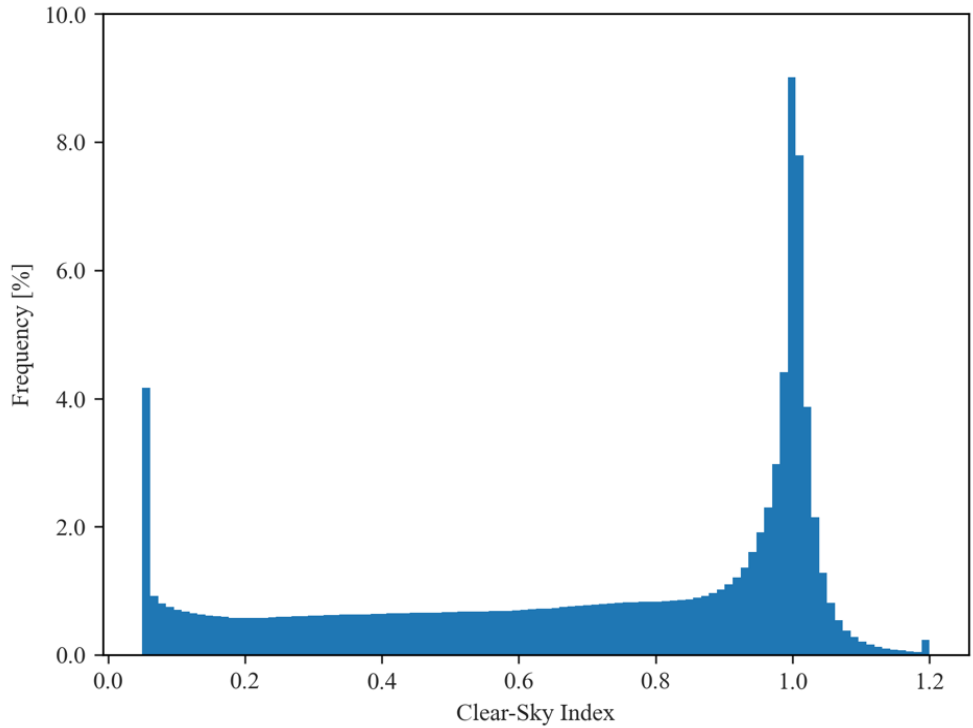

**Figure A6.** Frequency distribution of CSI pixels values over 2018 daytimes. The distribution is bimodal with two peaks at 0.05 and 1.

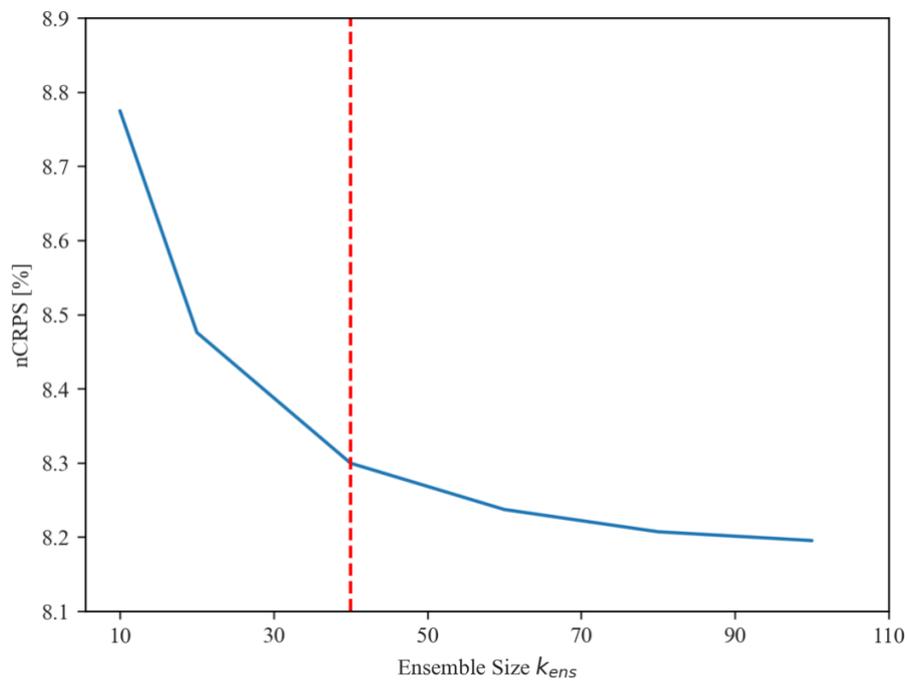

**Figure A7.** Normalized CRPS computed on the validation set. The loss strongly decreases up to approximately 40 members, which we selected as ensemble size for the forecast quality evaluation.

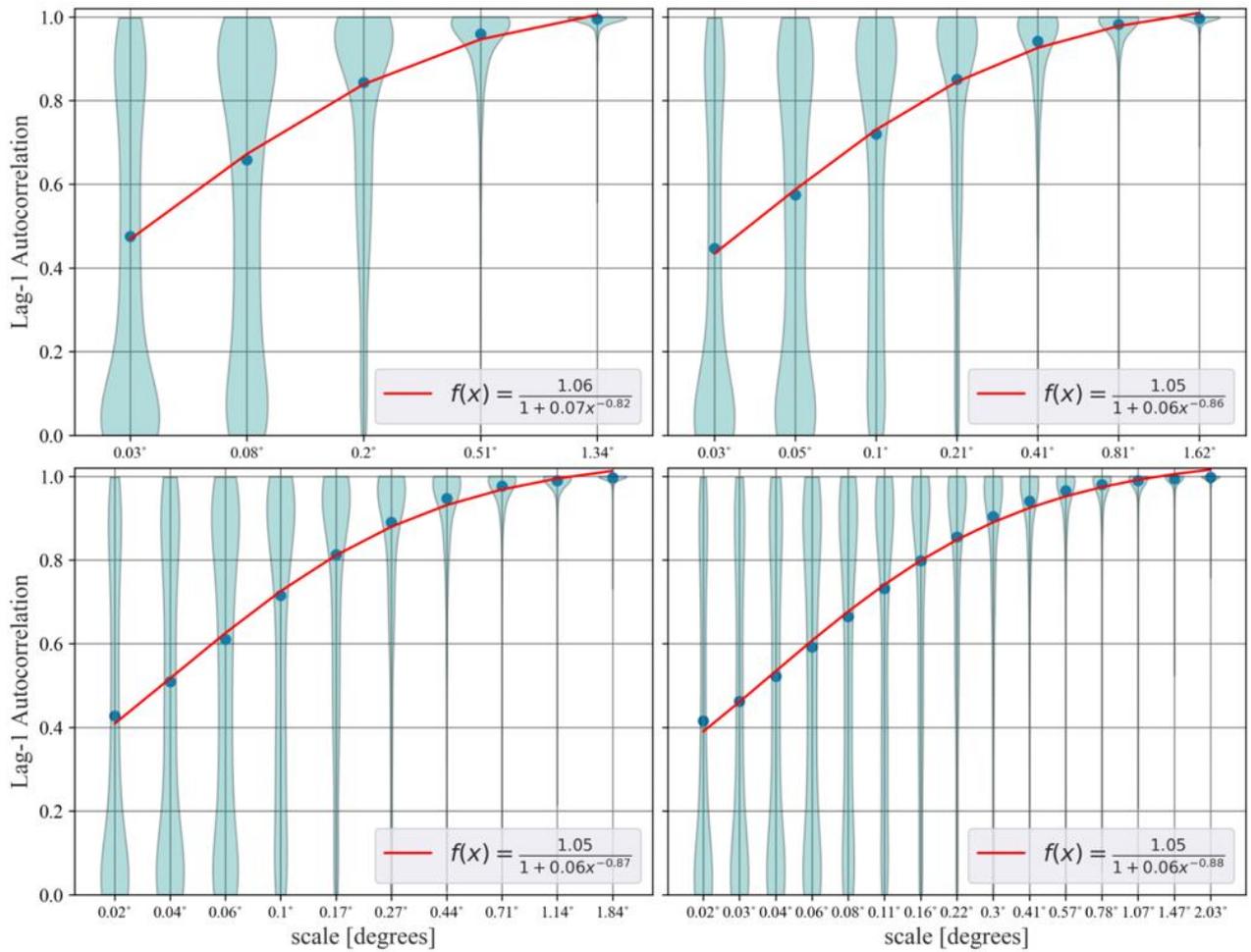

**Figure A8.** Kernel density estimates of the lag-1 autocorrelation for different decompositions of the CSI fields. The validation set is used to compute the metrics. A reciprocal power-law function is used to fit the means of the autocorrelation values for each cascade level. Different numbers of cascade levels are assessed: All of them confirm the scale-dependence of the CSI temporal variability and the function coefficients which do not significantly change with the number of cascade levels.

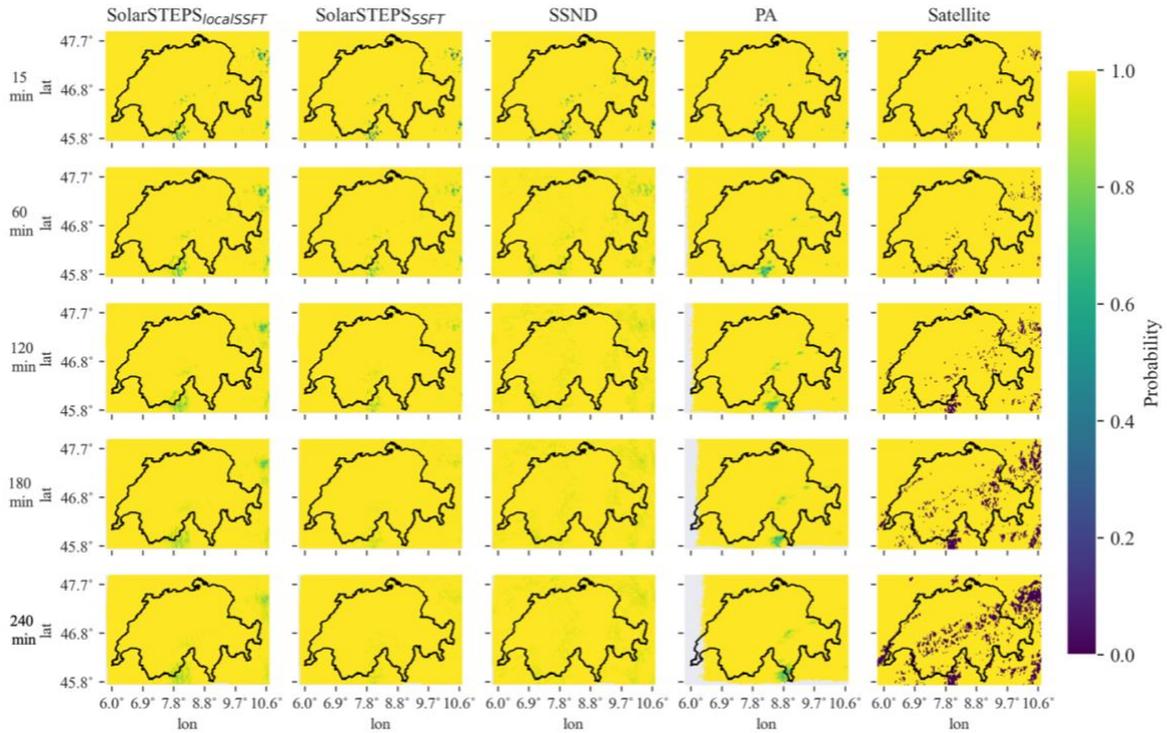

**Figure A9.** Probabilistic forecast in a sunny weather situation. 40 members are employed for each model to forecast the clear-sky index maps over Switzerland for the period ranging from 2018-07-09 07:30 am to 2018-07-09 11:15 am. The forecasts' pixels values are the probabilities of exceeding a threshold, which is set to 0.9 in this case. The date is chosen as it is the period with the lowest variability in the test set. The variability is computed as the standard deviation of the input maps. In this case, the input maps are the CSI maps ranging from 2018-07-09 06:30 am to 2018-07-09 7:15 am.

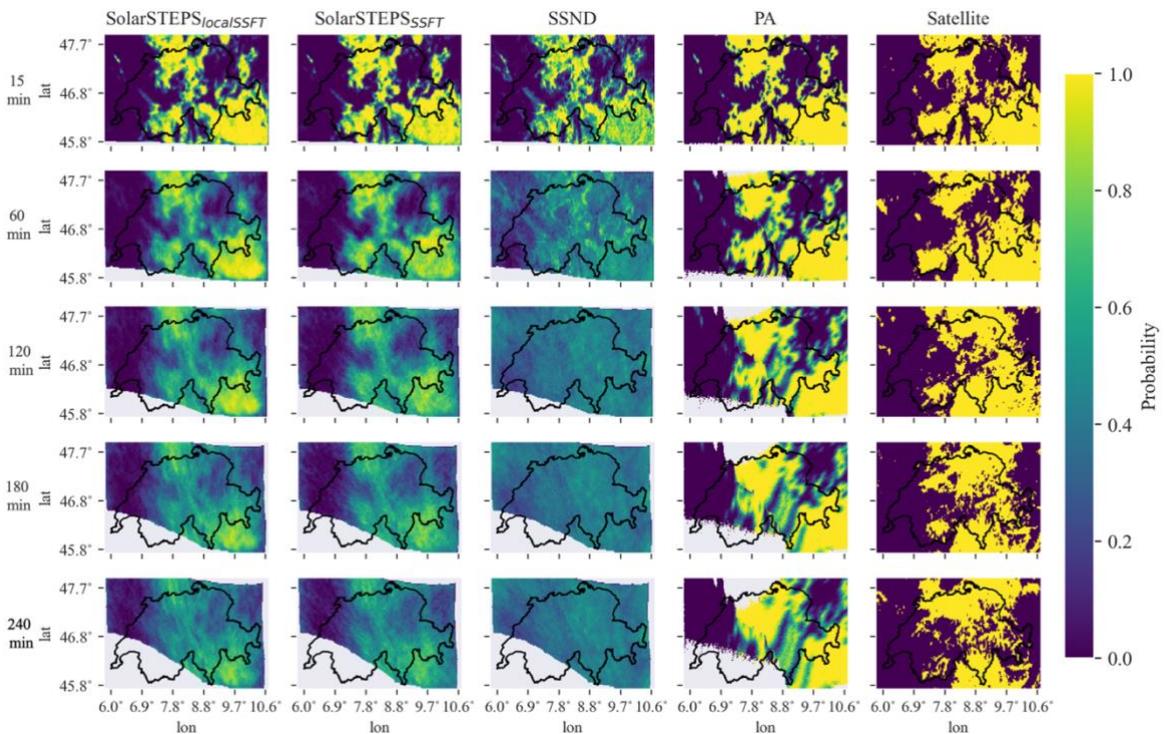

**Figure A10.** Probabilistic forecast in a variable weather situation. 40 members are employed for each model to forecast the clear-sky index maps over Switzerland for the period ranging from 2018-07-09 07:00 am to 2018-07-09 10:45 am. The forecasts' pixels values are the probabilities of exceeding a threshold, which is set to 0.9 in this case. The date is chosen as it is the period with the highest variability in the test set. The variability is computed as the standard deviation of the input maps. In this case, the input maps are the CSI maps ranging from 2018-10-07 06:00 am to 2018-07-09 6:45 am.